\documentstyle[12pt]{article}
\begin{document}
\begin{titlepage}
\title{Models of Supersymmetric \\ $U(2)\times U(1)$ Flavor Symmetry}
\author{Galit Eyal \\
 \small\it{Department of Particle Physics, Weizmann Institute of
 Science, Rehovot 76100, Israel}}
\date{\small{WIS-98/20/JULY-DPP}}
\maketitle

We use a $U(2)\times U(1)$ horizontal symmetry in order to construct 
supersymmetric models where the flavor structure of both quarks and
leptons is induced naturally. The supersymmetric flavor changing neutral
currents problem is solved by the degeneracy between sfermions induced by
the $U(2)$ symmetry. The additional $U(1)$ enables the generation of mass
ratios that cannot be generated by $U(2)$ alone. The resulting
phenomenology differs from that of models with either abelian or
$U(2)\times GUT$ symmetries. Our models give rise to interesting neutrino
spectra, which can incorporate the Super-Kamiokande results regarding
atmospheric neutrinos.
\end{titlepage}

\section{Introduction}
Approximate horizontal symmetries, $H$, can naturally explain the observed
flavor structure of fermions. With abelian 
symmetries~\cite{a:lns93}-\cite{a:hlept}
all mass ratios
and mixing angles are explained in a straightforward way. On the other
hand, with a $U(2)$ symmetry~\cite{a:pomarol1} it is quite difficult to
explain the large $m_t/m_b$ ratio and the different hierarchies in the 
down and up sectors. In order to overcome these problems, the $U(2)$ 
symmetry is often combined with a Grand Unified Theory (GUT) and a very
specific choice of flavon
representations~\cite{a:barb1}-\cite{a:tanimoto}.

Within Supersymmetry (SUSY), the horizontal symmetries should also 
suppress new contributions to Flavor Changing Neutral Currents (FCNC). In
models with a $U(2)$ symmetry and generations in ${\bf 2+1}$
representations (as in other models of non-abelian horizontal 
symmetry~\cite{a:dlk93}-\cite{a:bb96}),
the SUSY FCNC problem is
automatically solved by degeneracy between the first two sfermion
generations. On the other hand, with Abelian symmetries, the simple
alignment mechanism~\cite{a:nirnsei} does not give strong enough
suppression of the FCNC. In order to solve this problem, one is usually
led to rather specific $H$-charge assignments that yield a very precise
alignment.

We construct models with a $U(2)\times U(1)$ symmetry that combine the
advantages of the two frameworks. The $U(2)$ symmetry solves the SUSY FCNC
problem without alignment and the $U(1)$ symmetry accounts for the various
mass ratios without invoking a specific GUT structure. The resulting
phenomenology is different from that of either framework.

We assume that at some high-energy scale the symmetry $H\equiv U(2) \times
U(1)=SU(2)\times U(1)_{1}\times U(1)_{2}$ is realized. The symmetry
is broken in the following hierarchical way:
\begin{equation}
U(2)\times U(1) \stackrel{\epsilon}{\rightarrow} U(1)' \stackrel{\epsilon
'}{\rightarrow} 0.
\end{equation}
This is done by giving Vacuum Expectation Values (VEVs) to flavon fields.
We quantify all the breaking parameters as powers of a small parameter
$\lambda$ which we take to be of $O(0.2)$:
\begin{equation}
\epsilon \sim \lambda, \hspace{2cm} \epsilon ' \sim \lambda^{2}.
\end{equation}

The three generations are in {\bf{2+1}} representations. In difference 
with previous models of $U(2)$, we allow the different SM representations 
to carry different charges under $U(1)_{1}\times U(1)_{2}$. On one hand,
the model is not compatible with an underlying GUT, and loses some of the 
predictive power found in the GUT scenario. On the other hand, with a
small number of flavon fields, we are able to reproduce the mass matrices
of the quarks and the leptons and the CKM matrix elements, without
invoking any additional different mechanisms or symmetries in order to
solve specific problems. Within our framework the large ratio
$m_{t}/m_{b}$ can be explained without imposing large $\tan\beta$. Also,
the $\mu$ term can naturally be of order of the SUSY breaking scale, and
does not need to be put by hand.

Our model reproduces the following high-energy scale mass ratios:
\begin{eqnarray}
(m_{u};m_{c};m_{t})&\rightarrow& (\lambda^{7};\lambda^{4};1),\\
(m_{d};m_{s};m_{b})&\rightarrow& (\lambda^{7};\lambda^{5};\lambda^{3}),\\
(m_{e};m_{\mu};m_{\tau})&\rightarrow&
(\lambda^{8};\lambda^{5};\lambda^{3}).
\end{eqnarray}
The CKM elements are of the experimentally measured order of magnitude,
while the Kobayashi-Maskawa phase, $\delta_{KM}$, can receive any
experimentally allowed value, and is not restricted to be of $O(1)$. The
charges of the matter fields are chosen in such a way that the $U(1)'$
symmetry acts only on the first generation.

Due to the symmetry between the first two generations, a degeneracy 
between the corresponding sfermions is produced. The degeneracy can be
made to be very strong, but here we choose it to be of $O(\epsilon^{2})$. 
This degeneracy is strong enough to solve the SUSY FCNC problem, while 
mild enough to still allow approximate CP - a description of all CP
violating phenomena with small CP violating phases~\cite{a:mine,a:babu}.
Since we do not use the alignment mechanism, we do not necessarily
reproduce some of its generic features. In particular, $\Delta m_{D}$ is
not close to the experimental bound.

While in our model there is room for approximate CP as a solution to the
SUSY CP problem, we do not treat the CP violating phases explicitly. 
Within this framework there is also the possibility of relaxing the SUSY
CP problem by increasing the degeneracy between the first two sfermion
generations.

Our models allow various interesting structures of the neutrino sector. In
light of the recent results announced by Super-Kamiokande~\cite{a:superK}
which, in the three generations framework, imply
\begin{equation}
\Delta m_{23}^{2}\sim 5\times 10^{-3} \;eV^{2}, \hspace{2cm}
\sin^{2}2\theta_{23}\sim 1,
\end{equation}
we present two models for the structure of the lepton sector:

\begin{itemize}
\item lepton-model I: hierarchy $\hspace{2cm}$ $m_{\nu_{e}} \ll
m_{\nu_{\mu}} < m_{\nu_{\tau}}$,
\item lepton-model II: quasi-degeneracy $\hspace{0.5cm}$ $m_{\nu_{e}}\ll
|m_{\nu_{\mu}}| 
\simeq |m_{\nu_{\tau}}|$.
\end{itemize}

The structure of the paper is as follows. In section 2 we present the
quark sector of the model. Here we introduce all the flavon fields used in
both the quark and the lepton sectors. We study the implications of this
model to FCNC processes. In section 3 we present two extensions of the
model that describe the lepton sector. Our conclusions are summarized in
section 4.

\section{The quark sector}
The superfields of the quark and Higgs sectors of the Supersymmetric
Standard Model (SSM) carry $H$-charges, shown in table~\ref{t:hquarks}.
\begin{table}[hbct]
\begin{center}
\begin{tabular}{|c|c|c|}  
\hline
Field & $SU(2)$ & $(U(1)_{1},U(1)_{2})$ \\
\hline
$\phi_{u}$ & 1 & (0,0)\\
$\phi_{d}$ & 1 & (0,-2)\\
$\left( \begin{array}{c}
               Q_{1}\\   
               Q_{2}
               \end{array}\right)$ & 2 & (1,4)\\
$Q_{3}$ & 1 & (0,0) \\
$\left( \begin{array}{c}
               \bar{u}_{1}\\
               \bar{u}_{2}
               \end{array}\right)$ & 2 & (3,0)\\        
$\bar{u}_{3}$ & 1 & (0,0) \\
$\left( \begin{array}{c}
               \bar{d}_{1}\\
               \bar{d}_{2}
               \end{array}\right)$ & 2 & (3,0) \\
$\bar{d}_{3}$ & 1 & (0,6)  \\
\hline
\end{tabular}
\end{center}
\caption{$H$ charges of the Higgs and Quark superfields.}
\label{t:hquarks}
\end{table}
There, $Q_{i}$ are the quark doublets, $\bar{d}_{i}$ and $\bar{u}_{i}$ are
the down and up quark singlets, and $\phi_{u}$ and $\phi_{d}$ are the
Higgs doublet fields. The electroweak symmetry is spontaneously broken by
the VEVs of $\phi_{d}$ and $\phi_{u}$, and we assume that 
\begin{equation}
\tan\beta\equiv
\frac{\bigl<\phi_{u}\bigr>}{\bigl<\phi_{d}\bigr>}\sim\frac{1}{\lambda}.
\end{equation}
In addition we have standard model singlet superfields: two $U(2)$
doublets and two $U(2)$ singlets. Their $H$-charges are shown in 
table~\ref{t:hsinglet}.
\begin{table}[hbct]
\begin{center}
\begin{tabular}{|c|c|c|}  
\hline
Field & $SU(2)$ & $(U(1)_{1},U(1)_{2})$\\
\hline
$\phi_{1} =\left( \begin{array}{c}
               \phi_{11}\\
               \phi_{12}
               \end{array}\right)$ & $\bar{2}$ & (-1,-1)\\
$\phi_{2} =\left( \begin{array}{c}
               \phi_{21}\\
               \phi_{22} 
               \end{array}\right)$ & $\bar{2}$ & (-1,-2)\\
$\chi_{1}$ & 1 & (0,-2) \\
$\chi_{2}$ & 1 & (-2,0) \\
\hline
\end{tabular}
\end{center}
\caption{$H$ charges of the SM-singlet superfields.}
\label{t:hsinglet}
\end{table}
The horizontal symmetry is broken when some of the SM-singlet fields 
assume VEVs. We take for the VEVs:
\begin{eqnarray}
\frac{1}{M}\left( \begin{array}{c}
               \bigl<\phi_{11}\bigr>\\
               \bigl<\phi_{12}\bigr>
               \end{array}\right)& = \left( \begin{array}{c}
               \epsilon '\\
               \epsilon
               \end{array}\right)& \sim \left( \begin{array}{c}
               \lambda^{2}\\
               \lambda
               \end{array}\right) \\
\frac{1}{M}\left( \begin{array}{c}
               \bigl<\phi_{21}\bigr>\\
               \bigl<\phi_{22}\bigr>
               \end{array}\right)& \sim \left( \begin{array}{c}
               0\\
               \epsilon  
               \end{array}\right) & \sim \left( \begin{array}{c}
               0\\
               \lambda 
               \end{array}\right)\\
\frac{1}{M}\bigl<\chi_{1}\bigr>& \sim \epsilon  & \sim \lambda\\
\frac{1}{M}\bigl<\chi_{2}\bigr>& \sim \epsilon ' & \sim \lambda^{2}
\end{eqnarray}
where M is a scale in which the information about this breaking is
communicated to the SSM. The symmetries of the model allow the choice
$\bigl<\phi_{21}\bigr>=0$ and $\bigl<\phi_{22}\bigr>$,
$\bigl<\chi_{1}\bigr>$, $\bigl<\chi_{2}\bigr>$ real. The additional fields
will, in general, receive complex VEVs. 

The VEVs and charges allow us to estimate the quark mass matrices $M^{f}$
and the squark mass-squared matrices $\tilde{M}^{f^{2}}$. All the terms
allowed by SUSY and $U(2)\times U(1)$ are assumed to appear with 
coefficients of $O(1)$. When a parameter appears explicitly, it is assumed
to be the $O(1)$ coefficient of the corresponding term. We write the 
effective matrices derived after rotations needed to bring the kinetic 
terms into their canonical form~\cite{a:sequel,a:dps95,a:blr96}. We get:
\begin{equation}
M^{d}\sim 
\bigl<\phi_{d}\bigr>\left(\begin{array}{ccc}
\epsilon '^{3} & \epsilon '^{2}\epsilon & \epsilon '\epsilon^{5} \\
\epsilon '^{2}\epsilon & \epsilon '\epsilon^{2} & \epsilon^{4} \\
\epsilon '^{2}\epsilon^{3} & \epsilon '\epsilon^{4} & \epsilon^{2}
\end{array} \right),
\end{equation}
\begin{equation}
M^{u}\sim \bigl<\phi_{u}\bigr>\left( \begin{array}{ccc}
\epsilon '^{3}\epsilon & \epsilon '^{2}\epsilon^{2} & \epsilon '
\epsilon^{3} \\
\epsilon '^{2}\epsilon^{2} & \epsilon '\epsilon^{2} & \epsilon^{2}\\
\epsilon '^{2}\epsilon^{2} & \epsilon '\epsilon^{2} & 1
\end{array} \right),
\end{equation}
\begin{equation}
\tilde{M}^{q^{2}}_{LL}\sim \tilde{m}^{2}
\left( \begin{array}{ccc}
a+\epsilon^{2} & \epsilon '\epsilon & \epsilon '\epsilon^{4} \\
\epsilon '\epsilon & a+\epsilon^{2} & \epsilon^{2} \\
\epsilon '\epsilon^{4} & \epsilon^{2} & 1
\end{array} \right)
\end{equation}
($a$ is, as explained above, a coefficient of $O(1)$ that marks the 
degeneracy of the $\{11\}$ and $\{22\}$ entries, and $\tilde{m}$ is the
SUSY breaking scale),
\begin{equation}
\tilde{M}^{d^{2}}_{RR}\sim
\tilde{m}^{2}\left( \begin{array}{ccc}
a+\epsilon^{2} & \epsilon '\epsilon & \epsilon '^{2}\epsilon^{3} \\
\epsilon '\epsilon & a+\epsilon^{2} & \epsilon '\epsilon^{5} \\
\epsilon '^{2}\epsilon^{3} & \epsilon '\epsilon^{5} & 1
\end{array} \right),
\end{equation}
\begin{equation}
\tilde{M}^{u^{2}}_{RR}\sim
\tilde{m}^{2}\left( \begin{array}{ccc}
a+\epsilon^{2} & \epsilon '\epsilon & \epsilon '^{2}\epsilon^{2} \\
\epsilon '\epsilon & a+\epsilon^{2} & \epsilon '\epsilon^{2} \\
\epsilon '^{2}\epsilon^{2} & \epsilon '\epsilon^{2} & 1
\end{array} \right),
\end{equation}
\begin{equation}
\tilde{M}^{q^{2}}_{LR}\sim \tilde{m}M^{q}.
\end{equation}
Note that the ratio $m_{t}/m_{b}$ is explained by the horizontal 
symmetries, as is the difference in hierarchies between the up and the 
down sectors. This is in contrast to models where the structure of the
mass matrices is dictated by $U(2)$ alone. 

We can also estimate the size of the bilinear $\mu$ and B terms:
\begin{eqnarray}
\mu\sim\tilde{m}\epsilon,\\
m^{2}_{12}\sim\tilde{m}^{2}\epsilon.
\end{eqnarray}
Thus the horizontal symmetry solves the $\mu$-problem in the way suggested
in~\cite{a:nir}.

From the mass matrices we can estimate the mixing angles in the CKM
matrix. We find:
\begin{equation}
|V_{us}|\sim\lambda, \hspace{1cm} |V_{ub}|\sim\lambda^{4}, \hspace{1cm} 
|V_{cb}|\sim\lambda^{2}, \hspace{1cm} |V_{td}|\sim\lambda^{3}.
\end{equation}

In order to compare quark-squark-gaugino mixing with the experimental
bounds presented in~\cite{a:masiero1} we use the formula given 
in~\cite{a:nirnsei}:
\begin{equation}
\delta_{MN}^{f}\sim (V_{M}^{f}\tilde{M}^{f^{2}}_{MN}V_{N}^{f^{\dagger}})
/\tilde{m}^{2} \label{eq:delta}
\end{equation}
where $\{M,N\}=\{L,R\}$, and $V^{f}_{M}$ are the diagonalizing matrices of
$M^{f}$. The dimensionless $\delta^{q}_{MN}$ matrices have the simple
meaning of squark mass-squared matrices (normalized to the average squark
mass-squared $\tilde{m}^{2}$) in the basis where gluino couplings are
diagonal and quark mass matrices are diagonal. The comparison is
summarized in table~\ref{t:bquarks}. There, the phenomenological bounds
scale like $(\tilde{m}/1\; TeV)^{2}$, and the $CP$ violating phases are
assumed to be of $O(1)$.
\begin{table}[hbct]
\begin{center}
\begin{tabular}{|c|c|c|c|}
\hline
 & Process & Bound & Model\\
\hline
$Re(\delta_{12}^{d})^{2}_{LL}$ & $\Delta m_{K}$ & $\lambda^{3}$
& $\lambda^{6}$\\
$Re(\delta_{12}^{d})_{LL}(\delta_{12}^{d})_{RR}$ & $\Delta m_{K}$ &
$\lambda^{6}-\lambda^{7}$ & $\lambda^{6}$\\
$Re(\delta_{12}^{d})^{2}_{RR}$ & $\Delta m_{K}$ & $\lambda^{3}$ &
$\lambda^{6}$\\
$Re(\delta_{13}^{d})^{2}_{LL}$ & $\Delta m_{B}$ & $\lambda^{2}$ &
$\lambda^{6}$\\
$Re(\delta_{13}^{d})_{LL}(\delta_{13}^{d})_{RR}$ & $\Delta m_{B}$ &
$\lambda^{4}$ & $\lambda^{8}$\\
$Re(\delta_{13}^{d})^{2}_{RR}$ & $\Delta m_{B}$ & $\lambda^{2}$ &
$\lambda^{10}$\\
$Re(\delta_{12}^{u})^{2}_{LL}$ & $\Delta m_{D}$ & $\lambda^{2}$ &
$\lambda^{6}$\\
$Re(\delta_{12}^{u})_{LL}(\delta_{12}^{u})_{RR}$ & $\Delta m_{D}$ &
$\lambda^{4}$ & $\lambda^{6}$\\
$Re(\delta_{12}^{u})^{2}_{RR}$ & $\Delta m_{D}$ & $\lambda^{2}$ &
$\lambda^{6}$\\
$Im(\delta_{12}^{d})^{2}_{LL}$ & $\epsilon_{K}$ & $\lambda^{6}$ &
$\lambda^{6}$\\
$Im(\delta_{12}^{d})_{LL}(\delta_{12}^{d})_{RR}$ & $\epsilon_{K}$ &
$\lambda^{9}-\lambda^{10}$ & $\lambda^{6}$\\
$Im(\delta_{12}^{d})^{2}_{RR}$ & $\epsilon_{K}$ & $\lambda^{6}$ &
$\lambda^{6}$\\
\hline
\end{tabular}
\end{center}
\caption{Squark mass parameters: model predictions vs. phenomenological
bounds.}
\label{t:bquarks}
\end{table}
We learn the following points from table~\ref{t:bquarks}:
\begin{itemize}
\item $\Delta m_{K}$ receives SUSY contributions comparable to the SM
 ones.
\item Contrary to $U(2)\times GUT$ symmetry models~\cite{a:barb4}, SUSY
 contributions to $\Delta m_{B}$ are negligible compared to the SM ones. 
\item Contrary to abelian horizontal symmetry 
 models~\cite{a:sequel,a:nirnsei}, $\Delta m_{D}$ is not expected to be at
 the experimental limit, but rather $1-2$ orders of magnitude smaller.
\item In order not to exceed the measured value of $\epsilon_{K}$, the CP
 violating phase contributing to $Im(\delta^{d}_{12})_{LL}
 (\delta^{d}_{12})_{RR}$ should be small.
\end{itemize}

\section{The lepton sector}
Various anomalies in neutrino experiments provide further input to flavor
models~\cite{a:rasil}-\cite{a:bhssw98}.
In the following we show how either an hierarchical spectrum or
quasi-degenerate neutrinos consistent with the recent measurements of
atmospheric neutrinos~\cite{a:superK} are produced naturally in an 
extension of our model to the lepton sector.

\subsection{Model I: Hierarchy}
The $H$ charges of the lepton superfields in our model I are given in 
table~\ref{t:hlepton1}. There, $L_{i}$ are the lepton doublets, 
$\bar{l}_{i}$ and $\bar{s}_{i}$ are the charged lepton and neutrino
singlets. 
\begin{table}[hbct]
\begin{center}
\begin{tabular}{|c|c|c|}  
\hline
Field & $SU(2)$ & $(U(1)_{1},U(1)_{2})$\\
\hline
$\left( \begin{array}{c}
               L_{1}\\
               L_{2}
               \end{array}\right)$ & 2 & (1,0)\\
$L_{3}$ & 1 & (0,0)  \\
$\left( \begin{array}{c}
               \bar{l}_{1}\\
               \bar{l}_{2}
               \end{array}\right)$ & 2 & (3,6)\\
$\bar{l}_{3}$ & 1 & (0,6)  \\
$\left( \begin{array}{c}
               \bar{s}_{1}\\
               \bar{s}_{2}
               \end{array}\right)$ & 2 & (1,2)\\
$\bar{s}_{3}$ & 1 & (0,3)  \\
\hline
\end{tabular}
\end{center}
\caption{Model I: $H$ charges of the lepton superfields.}
\label{t:hlepton1}
\end{table}
We get:
\begin{equation}
M^{l}\sim \bigl<\phi_{d}\bigr>\left( \begin{array}{ccc}
\epsilon '^{3}\epsilon & \epsilon '^{2}\epsilon^{2} & \epsilon ' 
\epsilon^{3}\\
\epsilon '^{2}\epsilon^{2} & \epsilon '\epsilon^{2} & \epsilon^{2} \\
\epsilon '^{2}\epsilon^{3} & \epsilon '\epsilon^{2} & \epsilon^{2}
\end{array} \right).
\end{equation}
The matrices  $M^{\nu}_{RR}$ and $M^{\nu}_{LR}$ are given in their naive
form before the rotations to the canonical form are made:
\begin{equation}
M^{\nu}_{RR}\sim M_{L}\left( \begin{array}{ccc}
a\epsilon '^{2}\epsilon & a\epsilon '\epsilon^{2} & c\epsilon ' 
\epsilon^{2}\\
a\epsilon '\epsilon^{2} & b\epsilon^{2} & c\epsilon^{3}  \\
c\epsilon '\epsilon^{2}  & c\epsilon^{3} & \epsilon^{3}
\end{array} \right),
\end{equation}
\begin{equation}
M^{\nu}_{LR}\sim \bigl<\phi_{u}\bigr> \left( \begin{array}{ccc}
a\epsilon '^{2}  & (a+b)\epsilon '\epsilon & c\epsilon '\epsilon \\
(a-b)\epsilon '\epsilon & a\epsilon^{2} & c\epsilon^{2}\\
0 & d\epsilon  & 0
\end{array} \right).
\end{equation}
Using the see-saw mechanism, and arranging in the canonical form:
\begin{equation}
M^{\nu}_{LL}\sim  M^{\nu}_{LR}M^{\nu\;\;-1}_{RR}M^{\nu\;\;T}_{LR}
\sim \frac{\bigl<\phi_{u}\bigr>^{2}}{M_{L}}\left( \begin{array}{ccc}
\epsilon '^{2}\epsilon^{-1} & \epsilon ' & \epsilon ' \\
\epsilon ' & \epsilon & \epsilon \\
\epsilon ' & \epsilon & 1
\end{array} \right).
\end{equation}

The hierarchy of the neutrino masses in this model is:
\begin{equation}
m_{\nu_{e}} \ll m_{\nu_{\mu}}< m_{\nu_{\tau}}.
\end{equation}
There is no degeneracy between any of the neutrinos. Using as input the
new data from Super-Kamiokande~\cite{a:superK}, we find:
\begin{eqnarray}
m_{\nu_{e}} \sim \lambda^{3} m_{\nu_{\tau}} \sim 5\times 10^{-4} \;eV, 
\hspace{1cm}
&m_{\nu_{\mu}} \sim \lambda m_{\nu_{\tau}} \sim 0.01 \;eV, \hspace{1cm}
&m_{\nu_{\tau}} \sim 0.07 \;eV,\\
\sin\theta_{12}\sim\lambda, \hspace{1cm}
&\sin\theta_{13}\sim\lambda, \hspace{1cm}
&\sin\theta_{23}\sim 1.
\end{eqnarray}
We also get
\begin{equation}
M_{L}\sim 5\times 10^{14}\; GeV.
\end{equation}
The neutrinos do not contribute significantly to the dark matter. The mass
of $\nu_{\mu}$ together with the mixing angle $\sin\theta_{12}$ might
point at the large angle matter enhanced solution to the solar neutrino
problem (although the mass is a bit too large)~\cite{a:barger,a:bks98}.

For the slepton mass-squared matrices, we get
\begin{equation}
\tilde{M}^{l^{2}}_{LL}\sim \tilde{m}^{2}
\left( \begin{array}{ccc}
a+\epsilon^{2} & \epsilon '\epsilon & \epsilon '\epsilon^{2}\\
\epsilon '\epsilon & a+\epsilon^{2} & \epsilon^{2} \\
\epsilon '\epsilon^{2} & \epsilon^{2} & 1
\end{array} \right),
\end{equation}
\begin{equation}
\tilde{M}^{l^{2}}_{RR}\sim
\tilde{m}^{2}\left( \begin{array}{ccc}
a+\epsilon^{2} & \epsilon '\epsilon & \epsilon '^{2}\epsilon^{2} \\
\epsilon '\epsilon & a+\epsilon^{2} & \epsilon '\epsilon^{2} \\
\epsilon '^{2}\epsilon^{2} & \epsilon '\epsilon^{2} & 1
\end{array} \right),
\end{equation}
\begin{equation}
\tilde{M}^{s^{2}}_{RR}\sim
\tilde{m}^{2}\left( \begin{array}{ccc}
a+\epsilon^{2} & \epsilon '\epsilon & \epsilon '\epsilon \\
\epsilon '\epsilon & a+\epsilon^{2} & \epsilon^{2} \\
\epsilon '\epsilon & \epsilon^{2} & 1
\end{array} \right),
\end{equation}
\begin{equation}
\tilde{M}^{l^{2}}_{LR} \sim \tilde{m} M^{l}.
\end{equation}
The comparison between lepton-slepton-gaugino parameters, defined   
analogously to the definitions in the quark sector (eq.~\ref{eq:delta}),
and the experimental bounds presented in~\cite{a:masiero1}, is summarized
in table~\ref{t:bleptons}. There, the phenomenological bounds for the
process $\mu\rightarrow e\gamma$ scale like $(\tilde{m}/1\;TeV)^{2}$, 
while the bound for the Electric Dipole Moment (EDM) of the electron
scales like $(\tilde{m}/1 \;TeV)$. The $CP$ violating phases are assumed
to be of $O(1)$. The bounds appear only for processes for which the bound
is $\leq 1$. The following points should be noted in 
table~\ref{t:bleptons}:
\begin{itemize}
\item The decay $\mu \rightarrow e\gamma$, if the slepton masses are
 close to $m_{Z}$, is expected to be close to the experimental limit. 
\item The EDM can be close to the experimental limit, if the CP violating
 phases are large.
\end{itemize}

\subsection{Model II: Quasi-degeneracy}
The $H$ charges of the lepton superfields in our model II are given in
table~\ref{t:hlepton2}. We get: 
\begin{table}[hbct]
\begin{center}
\begin{tabular}{|c|c|c|}  
\hline
Field & $SU(2)$ & $(U(1)_{1},U(1)_{2})$\\
\hline
$\left( \begin{array}{c}
               L_{1}\\
               L_{2}
               \end{array}\right)$ & 2 & (3,1)\\
$L_{3}$ & 1 & (-2,1)  \\
$\left( \begin{array}{c}
               \bar{l}_{1}\\
               \bar{l}_{2}
               \end{array}\right)$ & 2 & (1,5)\\
$\bar{l}_{3}$ & 1 & (4,1)  \\
$\left( \begin{array}{c}
               \bar{s}_{1}\\
               \bar{s}_{2}
               \end{array}\right)$ & 2 & (3,1)\\
$\bar{s}_{3}$ & 1 & (-2,1)  \\
\hline
\end{tabular}
\end{center}
\caption{Model II: $H$ charges of the lepton superfields.}
\label{t:hlepton2}
\end{table}
\begin{equation}
M^{l}\sim \bigl<\phi_{d}\bigr>\left( \begin{array}{ccc}
\epsilon '^{3}\epsilon & \epsilon '^{2}\epsilon^{2}  & \epsilon '^{4}
\epsilon^{2} \\
\epsilon '^{2}\epsilon^{2} & \epsilon '\epsilon^{2} & \epsilon '^{3}
\epsilon^{2}\\
\epsilon '^{2}\epsilon^{4} & \epsilon '^{3}\epsilon^{2} & \epsilon '
\end{array} \right).
\end{equation}
The matrices  $M^{\nu}_{RR}$ and $M^{\nu}_{LR}$ have the following
structure: 
\begin{equation}
M^{\nu}_{RR}\sim M_{L}\left( \begin{array}{ccc}
a\epsilon '^{4} & a\epsilon '^{3}\epsilon & 0 \\
a\epsilon '^{3}\epsilon & a\epsilon '^{2}\epsilon^{2} & b\epsilon  \\
0  & b\epsilon & 0
\end{array} \right),
\end{equation}
\begin{equation}
M^{\nu}_{LR}\sim \bigl<\phi_{u}\bigr>\left( \begin{array}{ccc}
a\epsilon '^{4}  & (a+b)\epsilon '^{3}\epsilon & 0 \\
(a-b)\epsilon '^{3}\epsilon & a\epsilon '^{2}\epsilon^{2} & c\epsilon\\
0 & d\epsilon   & 0
\end{array} \right).
\end{equation}
Using the see-saw mechanism, and arranging in the canonical form, we get:
\begin{equation}
M^{\nu}_{LL}\sim\frac{\bigl<\phi_{u}\bigr>^{2}}{M_{L}}
\left( \begin{array}{ccc}
\epsilon '^{4} & \epsilon '^{3}\epsilon & \epsilon '\epsilon^{2} \\
\epsilon '^{3}\epsilon & \epsilon '^{2}\epsilon^{2} & \epsilon  \\
\epsilon '\epsilon^{2} & \epsilon & \epsilon '^{2}\epsilon^{3}
\end{array} \right).
\end{equation}

The hierarchy of the neutrino masses in this model is:
\begin{equation}
m_{\nu_{e}} \ll |m_{\nu_{\mu}}| \simeq |m_{\nu_{\tau}}|.
\end{equation}
The degeneracy between $m_{\nu_{\mu}}$ and $m_{\nu_{\tau}}$ is 
$O(\lambda^{5})$. Analyzing this using the new data from Super-Kamiokande,
we find:
\begin{eqnarray}
m_{\nu_{e}} \sim \lambda^{7}m_{\nu_{\tau}}, \hspace{1cm}
m_{\nu_{\mu}}\simeq m_{\nu_{\tau}} \sim 3 \;eV,\\
\sin\theta_{12} \sim\lambda^{2}, \hspace{1cm}
\sin\theta_{13}< \lambda^{3},\hspace{1cm}
\sin\theta_{23}\simeq \frac{1}{\sqrt{2}},
\end{eqnarray}  
and 
\begin{equation}
M_{L}\sim 2\times 10^{12} \;GeV.
\end{equation}
Here the neutrinos play an important role in structure formation and 
contribute a significant part to the hot dark matter. The spectrum,
however, does not seem to be compatible with any of the suggested
solutions to the solar neutrino problem~\cite{a:barger,a:bks98}.

The sfermion mass-matrices have the following structure:
\begin{equation}
\tilde{M}^{l^{2}}_{LL}\sim \tilde{m}^{2}
\left( \begin{array}{ccc}
a+\epsilon^{2} & \epsilon '\epsilon & \epsilon '^{3}\epsilon^{2}\\
\epsilon '\epsilon & a+\epsilon^{2} & \epsilon '^{2}\epsilon^{2} \\
\epsilon '^{3}\epsilon^{2} & \epsilon '^{2}\epsilon^{2} & 1
\end{array} \right),
\end{equation}
\begin{equation}
\tilde{M}^{l^{2}}_{RR}\sim
\tilde{m}^{2}\left( \begin{array}{ccc}
a+\epsilon^{2} & \epsilon '\epsilon & \epsilon '\epsilon^{4} \\
\epsilon '\epsilon & a+\epsilon^{2} & \epsilon '^{2}\epsilon^{2} \\
\epsilon '\epsilon^{4} & \epsilon '^{2}\epsilon^{2} & 1
\end{array} \right),
\end{equation}
\begin{equation}
\tilde{M}^{s^{2}}_{RR}\sim
\tilde{m}^{2}\left( \begin{array}{ccc}
a+\epsilon^{2} & \epsilon '\epsilon & \epsilon '^{3}\epsilon^{2} \\
\epsilon '\epsilon & a+\epsilon^{2} & \epsilon '^{2}\epsilon^{2} \\
\epsilon '^{3}\epsilon^{2} & \epsilon '^{2}\epsilon^{2} & 1
\end{array} \right).
\end{equation}
The comparison between lepton-slepton-gaugino parameters and the 
experimental bounds presented in~\cite{a:masiero1}, is summarized in
table~\ref{t:bleptons}. 
\begin{table}[hbct]
\begin{center}
\begin{tabular}{|c|c|c|c|c|}
\hline
& Process & Bound & Model I & Model II\\
\hline
$|(\delta_{12}^{l})_{LL}|$ & $\mu\rightarrow e\gamma$ & 1 & $\lambda^{2}$
& $\lambda^{3}$\\
$|(\delta_{12}^{l})_{RR}|$ & $\mu\rightarrow e\gamma$ & 1 & $\lambda^{3}$
& $\lambda^{3}$\\
$|(\delta_{12}^{l})_{LR}|$ & $\mu\rightarrow e\gamma$ & $\lambda^{5}$ &
$\lambda^{6} \frac{\bigl<\phi_{d}\bigr>}{\tilde{m}}$ &
$\lambda^{6} \frac{\bigl<\phi_{d}\bigr>}{\tilde{m}}$\\
$|Im(\delta_{11}^{l})_{LR}|$ & EDM & $\lambda^{8}$ &
$\lambda^{7} \frac{\bigl<\phi_{d}\bigr>}{\tilde{m}}$ &
$\lambda^{7} \frac{\bigl<\phi_{d}\bigr>}{\tilde{m}}$\\
\hline
\end{tabular}
\end{center}
\caption{Slepton mass parameters: model predictions vs. phenomenological
bounds.}
\label{t:bleptons}
\end{table}
We point out that
\begin{itemize}
\item If the CP violating phases are large, the EDM can be close to the
 experimental limit.
\end{itemize}

\section{Conclusions}
Approximate flavor symmetries naturally explain the smallness and
hierarchy of the flavor parameters in SUSY models, while suppressing
sources for FCNC. Abelian horizontal symmetries explain the mass ratios in
a straightforward way, but need to invoke an alignment mechanism through
specific $H$ charge assignments in order to suppress FCNC. Horizontal
$U(2)$ symmetries suppress FCNC with a built-in degeneracy between the
first two sfermion generations, but need the framework of $GUT$ in order
to explain various mass ratios.

In this work, we presented a hybrid model of abelian and non-abelian
symmetries. The model combines the characteristics of both symmetries in
such a way as to produce all the required flavor parameters and
suppressions naturally, with no additional ingredient. The $U(2)$ symmetry
allows for a hierarchical breaking $U(2)\times U(1)
\stackrel{\epsilon}{\rightarrow} U(1)' \stackrel{\epsilon '}{\rightarrow}
0$ and gives the solution to the SUSY FCNC problem. The additional $U(1)$
enables generation of various mass ratios and mixing parameters in a 
simple way. It also allows for a natural solution of the SUSY $\mu$
problem.

The phenomenology of the hybrid model is different than that of either
abelian or non-abelian symmetry models. Unlike in usual non-abelian 
models, here different SM representations carry different charges so the
model is not compatible with GUT. On the other hand, $\Delta m_{D}$ is not
close to the experimental limit, as it is in models with alignment. This
framework leaves room for different possible solutions to the SUSY CP
problem, including approximate CP.

Different viable neutrino spectra can arise within this framework. We gave
two examples both of which are compatible with the recent observations of
atmospheric neutrinos by Super-Kamiokande. The first produces a hierarchy
of neutrino masses, while the second produces quasi-degenerate neutrinos,
that might play a significant role in cosmology.

The model presented here is not unique. It intends to demonstrate how
within the hybrid framework a simple model with very few flavon fields
can be built, that at the same time agrees with all measured flavor
parameters and suggests attractive spectra for the neutrino masses.\\

{\bf Acknowledgment}\\
 I thank Yossi Nir for many useful discussions and comments on the
manuscript.

{}



\begin{thebibliography}{99}
\bibitem{a:lns93} M. Leurer, Y. Nir and N. Seiberg, Nucl. Phys. B398
 (1993) 319, hep-ph/9212278.
\bibitem{a:sequel} M. Leurer, Y. Nir and N. Seiberg, Nucl. Phys. B420
 (1994) 468,  hep-ph/9310320.
\bibitem{a:dps95} E. Dudas, S. Pokorski and C.A. Savoy, Phys. Lett. B356
 (1995) 45, hep-ph/9504292. 
\bibitem{a:blr96} P. Binetruy, S. Lavignac and P. Ramond, Nucl. Phys. B477
 (1996) 353, hep-ph/9601243.
\bibitem{a:hlept} Y. Grossman and Y. Nir, Nucl. Phys. B448 (1995) 30,
 hep-ph/9502418. 
\bibitem{a:pomarol1} A. Pomarol and S. Tommasini, Nucl. Phys. B466
 (1996) 3, hep-ph/9507462.
\bibitem  {a:barb1} R. Barbieri, G. Dvali and L.J. Hall, Phys. Lett.
 B377 (1996) 76, hep-ph/9512388.
\bibitem{a:barb2} R. Barbieri and L.J. Hall, Nuovo Cim. A110 (1997) 1,
 hep-ph/9605224.
\bibitem{a:barb3} R. Barbieri, L.J. Hall, S. Raby and A. Romanino, Nucl.
 Phys. B493 (1997) 3, hep-ph/9610449.  
\bibitem{a:barb4} R. Barbieri, L.J. Hall and A. Romanino, Phys. Lett.
 B401 (1997) 47, hep-ph/9702315. 
\bibitem{a:carohall} C.D. Carone and  L.J. Hall, Phys. Rev. D56 (1997)
 4198, hep-ph/9702430. 
\bibitem{a:tanimoto} M. Tanimoto, Phys. Rev. D57 (1998) 1983,
 hep-ph/9706497.
\bibitem{a:dlk93} M. Dine, R. Leigh and A. Kagan, Phys. Rev. D48 (1993)
 4269, hep-ph/9304299.
\bibitem{a:ps93} P. Pouliot and N. Seiberg, Phys. Lett. B318 (1993) 169,
 hep-ph/9308363.
\bibitem{a:hm95} L.J. Hall and H. Murayama, Phys. Rev. Lett. 75 (1995) 
 3985, hep-ph/9508296.
\bibitem{a:chm96} C.D. Carone, L.J. Hall and H. Murayama, Phys. Rev. D54
 (1996) 2328, hep-ph/9602364.
\bibitem{a:bb96} K.S. Babu and S.M. Barr, Phys. Lett. B387
 (1996) 87, hep-ph/9606384.
\bibitem{a:nirnsei} Y. Nir and N. Seiberg, Phys. Lett. B309 (1993) 337,
 hep-ph/9304307.
\bibitem{a:mine} G. Eyal and Y. Nir, Hep-ph/9801411 (to be published in
 Nucl. Phys. B). 
\bibitem {a:babu} K.S. Babu and S.M. Barr. Phys. Rev. D49 (1994) R2156,
 hep-ph/9308217.
\bibitem{a:superK} Y. Fukuda et al. (The Super-Kamiokande
 Collaboration), hep-ex/9807003. 
\bibitem{a:nir} Y. Nir, Phys. Lett. B354 (1995) 107, hep-ph/9504312.   
\bibitem{a:masiero1} F. Gabbiani, E. Gabrielli, A. Masiero and L.
 Silvestrini, Nucl. Phys. B477 (1996) 321, hep-ph/9604387.
\bibitem{a:rasil} A. Rasin and J.P. Silva, Phys. Rev. D49 (1994) 20,
 hep-ph/9309240.
\bibitem{a:ps94} S.T. Petcov and A.Yu. Smirnov, Phys. Lett. B322 (1994)
 109, hep-ph/9311204.
\bibitem{a:calmo} D.O. Caldwell and R.N. Mohapatra, Phys. Rev. D50
 (1994) 3477, hep-ph/9402231. 
\bibitem{a:quasi} P. Binetruy, S. Lavignac, S. Petcov and P. Ramond,
 Nucl. Phys. B496 (1997) 3, hep-ph/9610481. 
\bibitem{a:fukutan} M. Fukugita, M. Tanimoto and T. Yanagida, Phys. Rev.
 D57 (1998) 4429, hep-ph/9709388.
\bibitem{a:casher} C.D. Carone and M. Sher, Phys. Lett. B420 (1998) 83,
 hep-ph/9711259.
\bibitem{a:irlara} N. Irges, S. Lavignac and P. Ramond, hep-ph/9802334.
\bibitem{a:barger} V. Barger, S. Pakvasa, T.J. Weiler and K. Whisnant,
 hep-ph/9806328.
\bibitem{a:elwirr} J.K. Elwood, N. Irges and P. Ramond, hep-ph/9807228.
\bibitem{a:bhssw98} R. Barbieri, L.J. Hall, D. Smith, A. Strumia and N.
 Weiner, hep-ph/9807235.
\bibitem{a:bks98} J.N. Bahcall, P.I. Krastev and A.Yu. Smirnov,
 hep-ph/9807216.
\end{thebibliography}
\end{document}